\documentclass[%
reprint,
superscriptaddress,
 amsmath,amssymb,
 aps,
 pra,
]{revtex4-1}
\usepackage[dvipsnames]{xcolor}
\definecolor{mblue}{RGB}{42, 54, 144} 
\usepackage{hyperref}
\hypersetup{backref,pdfpagemode=FullScreen,colorlinks=true,breaklinks,urlcolor=mblue,linkcolor=mblue,citecolor=mblue}
\usepackage{graphicx}
\usepackage{dcolumn}
\usepackage{bm}
\usepackage{physics}
\usepackage{braket}
\usepackage{amssymb}
\usepackage{changes}
\usepackage{pdfpages}
\makeatletter
\AtBeginDocument{\let\LS@rot\@undefined}
\makeatother

\begin{document}


\title{Extreme Spin Squeezing from Deep Reinforcement Learning}


\author{Feng Chen}
\affiliation{State Key Laboratory of Low-Dimensional Quantum Physics, Department of Physics, Tsinghua University, Beijing 100084, China}
\author{Jun-Jie Chen}
\affiliation{State Key Laboratory of Low-Dimensional Quantum Physics, Department of Physics, Tsinghua University, Beijing 100084, China}

\author{Ling-Na Wu}
\email{lnwu@pks.mpg.de}
\affiliation{Max Planck Institute for the Physics of Complex Systems, N\"{o}thnitzer Stra{\ss}e 38, 01187 Dresden, Germany}

\author{Yong-Chun Liu}
\email{ycliu@tsinghua.edu.cn}
\affiliation{State Key Laboratory of Low-Dimensional Quantum Physics, Department of Physics, Tsinghua University, Beijing 100084, China}
\affiliation{Frontier Science Center for Quantum Information, Beijing, China}
\affiliation{Collaborative Innovation Center of Quantum Matter, Beijing 100084, China}

\author{Li You}
\email{lyou@tsinghua.edu.cn}
\affiliation{State Key Laboratory of Low-Dimensional Quantum Physics, Department of Physics, Tsinghua University, Beijing 100084, China}
\affiliation{Frontier Science Center for Quantum Information, Beijing, China}
\affiliation{Collaborative Innovation Center of Quantum Matter, Beijing 100084, China}
\affiliation{Beijing Academy of Quantum Information Sciences, Beijing 100193, China}


\date{\today}

\begin{abstract}
Spin squeezing (SS) is a recognized resource for realizing measurement precision beyond the standard quantum limit $\propto 1/\sqrt{N}$.
The rudimentary one-axis twisting (OAT) interaction can facilitate SS and has been realized in diverse experiments, but it cannot achieve extreme SS for precision at Heisenberg limit $\propto 1/{N}$.
Aided by deep reinforcement learning (DRL), we discover size-independent universal rules for realizing nearly extreme SS with OAT interaction using merely a handful of rotation pulses. 
More specifically, only 6 pairs of pulses are required for up to $10^4$ particles, while the time taken to reach extreme SS remains on the same order of the optimal OAT squeezing time, which makes our scheme viable for experiments that reported OAT squeezing. 
This study highlights the potential of DRL for controlled quantum dynamics.
\end{abstract}


\maketitle


Squeezed spin state \cite{ueda1993, ramsey1, *ramsey1_2, ma2011} refers to a class of spin entangled state
whose uncertainty in one spin component perpendicular to the mean spin direction
is smaller than the classical limit $\sqrt{N}/2$ of a coherent spin state (CSS) of $N$ polarized (pseudo-) spin-1/2 particles ($\hbar =1 $ hereafter).
It can be applied in quantum metrology, e.g., through Ramsey interferometry \cite{ramsey1,ramsey2,ramsey3,ramsey5}, to reach precision beyond the standard quantum limit $\propto 1/\sqrt{N}$.
Extreme spin squeezing (SS) with precision at Heisenberg limit (HL) $\propto 1/N$ is realized from two-axis counter-twisting (TACT) interaction \cite{ueda1993}, although unfortunately TACT interaction does not arise easily in any known systems. 
Proposals for generating effective TACT interactions \cite{you2001,zxf2017,jzhu2017,tat1,tat2,tat4,tat6,xiao2017} are faced with challenges in their respective realizations.
An alternative approach to realize SS is based on one-axis twisting (OAT) interaction~\cite{ueda1993}, which occurs ubiquitously in systems of interacting spins, e.g., through large-detuned atom-photon coupling in an optical cavity \cite{oat2005,oat2010} or via atomic collisions in a Bose-Einstein condensate \cite{gross2010,riedel2010}.
However, SS from OAT interaction only scales as $1/N^{2/3}$ \cite{ueda1993}, which falls short of the HL of extreme squeezing $\propto 1/N$. 
Hence, developing approaches capable of improved SS based on OAT interaction constitutes a topical area of research.
A prominent model that has attracted widespread attention is to augment OAT interaction with a transverse coherent field \cite{ramsey4,law2001,jin2007,liu2011,guo2015,wln2015,duan2013}.
Promising proposals based on this model include, for instance, transforming OAT to TACT by using a periodic train of $\pm\pi/2$ spin rotation pulses \cite{liu2011}, or a periodically modulated transverse field \cite{guo2015}.
Their applications however call for rather stringent experimental conditions, with the former requiring a large number of high-precision pulses, while the latter demanding a high modulation frequency \cite{wln2015}.
An alternative proposal makes use of only a few optimized pulses along the mean spin direction \cite{duan2013}. 
While improvement over OAT is achieved, the evolution time required is greatly prolonged. 
It makes this scheme less robust against particle losses or decoherence. 
Despite intense efforts devoted over the years, it remains a difficult and unsolved task to find an experimentally feasible scheme implementable with state-of-the-art lab techniques.

One promising solution is to seek help from deep reinforcement learning (DRL), which has attracted much attention in recent years for its effective and flexible handling of dynamical systems. 
Without prior knowledge \cite{phase1}, DRL is capable of achieving an optimized policy and providing perspectives and solutions otherwise impossible or difficult to comprehend in optimal control.
The advantages of DRL over traditional theories and algorithms are already demonstrated with applications in quantum phase transition \cite{phase1}, quantum state preparation \cite{state1}, and quantum gate control \cite{dlzhou2019}, etc. 
However, DRL is computationally demanding and consumes enormous amount of resources for quantum systems with large degrees of freedom. 
For example, a typical DRL process requires thousands or even millions of training episodes, hence is prohibitive to directly handle systems with large particle number $N$.

In this Rapid Communication, aided by DRL, we propose a scheme applicable in large-sized systems for realizing extreme SS based on OAT interaction with a small number of rotation pulses. 
We first employ DRL in small-sized systems to find out size-independent universal rules for squeezing manipulation.
These rules are then applied to systems with large particle numbers. 
Limited computational resources are found to be sufficient even for quantum systems with large degrees of freedom. 
For $N=10^4$, we show that extreme SS close to the HL can be reached using only 6 pairs of rotation pulses applied at selected instants. 
Moreover, the evolution time required to reach optimal squeezing is on the same order of that for OAT, implicating that it is potentially robust against particle loss or decoherence.

We begin by describing our scheme applied to the model system of a condensate with pseudo-spin-$1/2$ atoms, whose Hamiltonian is given by \cite{gross2010,riedel2010}
\begin{equation}
\label{eq1}
H = \chi J_z^2 + \Omega(t)J_y.
\end{equation}
Here $J_{\mu} = \sum_k\sigma_{\mu}^{(k)}/2~(\mu=x,y,z)$ defines collective spin components with the Pauli matrices $\sigma_{x,y,z}^{(k)}$ for the $k$-th atom.
The Hamiltonian (\ref{eq1}) is composed of two terms: the first term $\chi J_z^2$ describes OAT interaction of strength $\chi$, which originates from, for example, spin exchange between atoms \cite{ma2011}; the second term $\Omega(t)J_y$ denotes the interaction between atoms and the external field with Rabi frequency $\Omega(t)$, which can be used to effect a spin rotation of angle $\theta= \int^{t_0+\Delta t}_{t_0}\Omega(t)dt$ along the $y$-axis over a short pulse duration $\Delta t$.
Starting from a CSS polarized along the $x$-axis with isotropic distributions for the transverse spin component fluctuations, the bare OAT interaction squeezes the transverse distribution and gives rise to a squeezed spin state with an optimal SS coefficient $\propto 1/N^{2/3}$ \cite{ueda1993}.
This study aims at reaching extreme squeezing, by applying the external field with controlled pulse sequence.
To make it experimentally feasible, the control pulse manipulations should be few and the evolution time should be short.

\begin{figure}
\begin{center}
\includegraphics[width=\columnwidth]{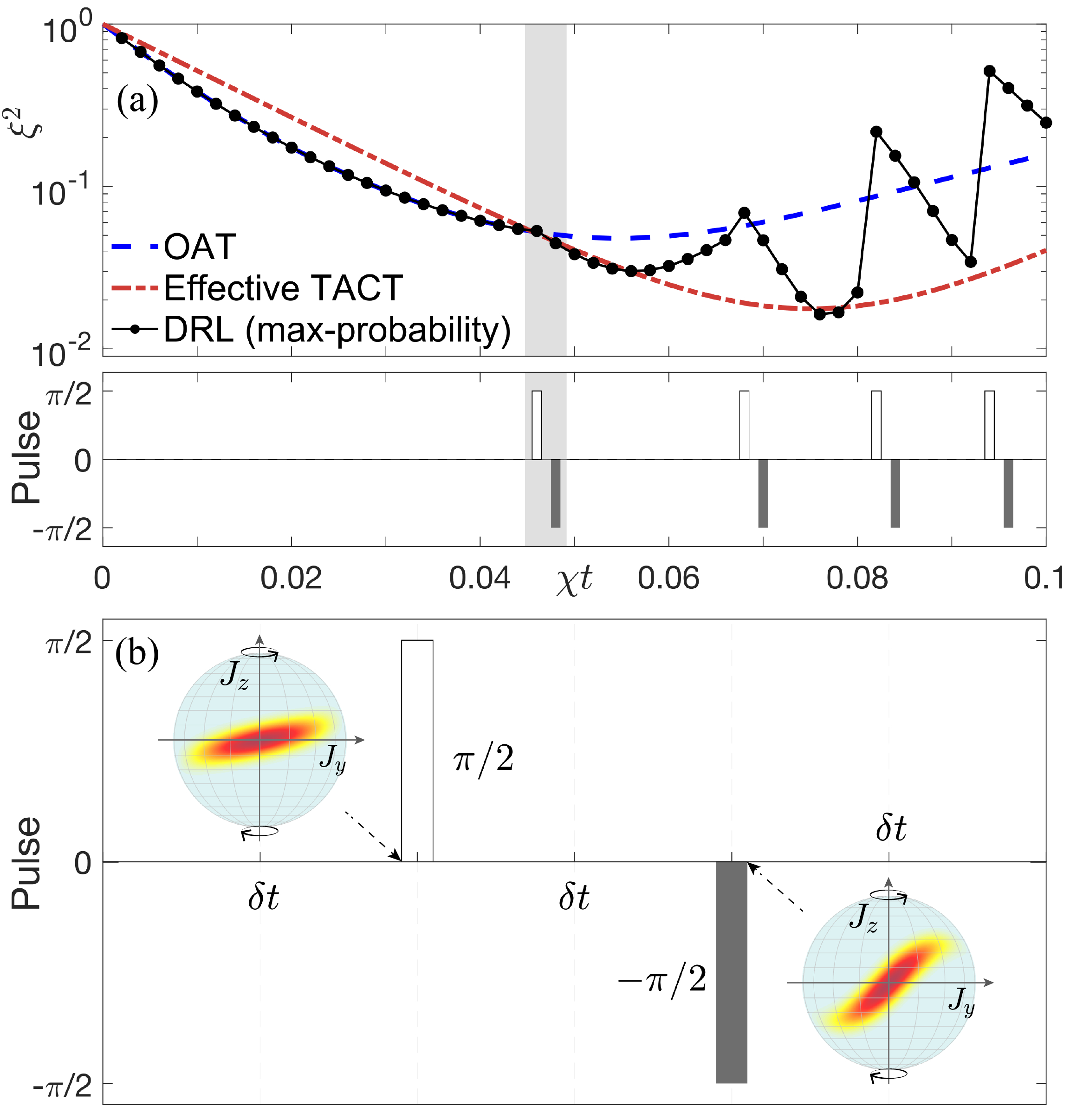}
\caption{(a) Evolution of SS coefficients are compared
among the max-probability protocol from DRL (black filled circles with solid guiding line) under Hamiltonian (\ref{eq1}), OAT (blue dashed line) with Hamiltonian $H_{\rm{OAT}} = \chi J_z^2$, and the effective TACT (red dash-dotted line) with Hamiltonian $H_{\rm{TACT}}=\chi(J_z^2-J_y^2)/3$ for $N = 100$. The initial state is a CSS polarized along $x$-axis.
The lower panel illustrates the corresponding DRL pulse sequence.
A zoomed-in region for one of the pairs (shaded) is shown in (b), with the distributions on the Bloch spheres denote the squeezed states before and after the $\pm\pi/2$ pulse pair, representing a nonlinear anti-clockwise rotation around the $x$-axis.}
\vspace{-0.7cm}
\label{fig1}
\end{center}
\end{figure}

First we employ DRL to obtain the pulse sequence for small-sized systems with $N\leq100$. 
The total training duration $t_c$ is set to be around the optimal squeezing time of the effective TACT (with reduced strength $\chi/3$) \cite{liu2011}. 
It is divided into $t_c/\delta t$ steps with fixed interval $\delta t$.
A typical DRL training is described as follows (more details can be found in the Supplementary Material \cite{supp2}). At each time step $t$, the agent observes environment (atoms here) through state $s_t \in \mathcal{S}$ and takes action $a_t \in \mathcal{A}$ guided by some policy $\varpi(a_t|s_t)$.
The environment then evolves to $s_{t+1}$ and returns a scalar reward $r_t \in \mathcal{R}$ back to the agent.
More specifically for our problem, the ``state'' ($\mathcal{S}$) contains five observables $s \in\{\expval{J_x},\expval{J_y},\expval{J_z},\expval{J_x^2},\expval{J_z^2}\}$.
The ``action'' ($\mathcal{A}$) consists of three discrete operations $a \in \{0,\pm\theta\}$, where $a=0$ means evolving for a time interval $\delta t$ under the bare OAT interaction, while $a=\pm\theta$ means a short rotation pulse with angle $\pm\theta$ applied before a $\delta t$ evolution under the OAT interaction.
The SS coefficient $\xi^2=4(\Delta\mathbf{J}_{\bot})_{\rm{min}}^2/N$ \cite{ueda1993} is employed as ``reward'' ($\mathcal{R}$), where $(\Delta\mathbf{J}_{\bot})_{\rm{min}}^2$ denotes the minimal fluctuation of the transverse spin component perpendicular to the mean spin direction.
To avoid sparse rewarding \cite{sparse1,sparse2} and to improve learning efficiency, we decompose the total achievable reward into a logarithm sum
\begin{equation}
\label{ }
R_{\rm{tot}}=\sum_{j=1}^nr_j=\sum_{j=1}^n\log_{10} \frac{\xi^2(t_{j-1})}{\xi^2(t_{j})},
\end{equation}
such that at each time step $t_j$, a reward $r_j$, the instantaneous decrease of SS coefficient, is fed back to the agent.
Proximal policy optimization (PPO) algorithm \cite{ppo2017} is employed to update the policy $\varpi(a|s)$ due to its stability and sophisticated control over the training processes.
After training, the PPO algorithm adopts an optimal stochastic policy which returns a distribution in action space $\varpi^*(a|s)$, satisfying $\sum_{a\in\mathcal{A}}\varpi^*(a|s) \equiv 1$ for a given $s$.
A specific protocol, i.e., pulse sequence, can be generated from the optimal policy $\varpi^*(a|s)$ in two ways.
One chooses the action (at each time step) $a = \max_{a'}\varpi^*(a'|s)$ with the maximum probability (max-probability for short)
and the other selects the best among multiple protocols generated from a sampling based on $\varpi^*(a|s)$ (with-selection  for short).
In actual DRL training tasks, a neural network is used to parameterize the actor and critic network employed in PPO.

Figure \ref{fig1}(a) shows the training result (max-probability) and its comparison with that of OAT and the effective TACT for $N = 100$.
Clearly, the result from DRL outperforms OAT, and the realized optimal SS approaches (or even outperforms) the effective TACT. 
Only two pairs of pulses are required for reaching the optimal SS, to be respectively applied at $\chi t\simeq 0.045$ and $\chi t\simeq 0.068$. 
The corresponding time for the optimal SS is $\chi t\simeq 0.075$, which is on the same order of the optimal OAT squeezing time ($\chi t \simeq 0.055$).

\begin{figure}
\begin{center}
\includegraphics[width=\columnwidth]{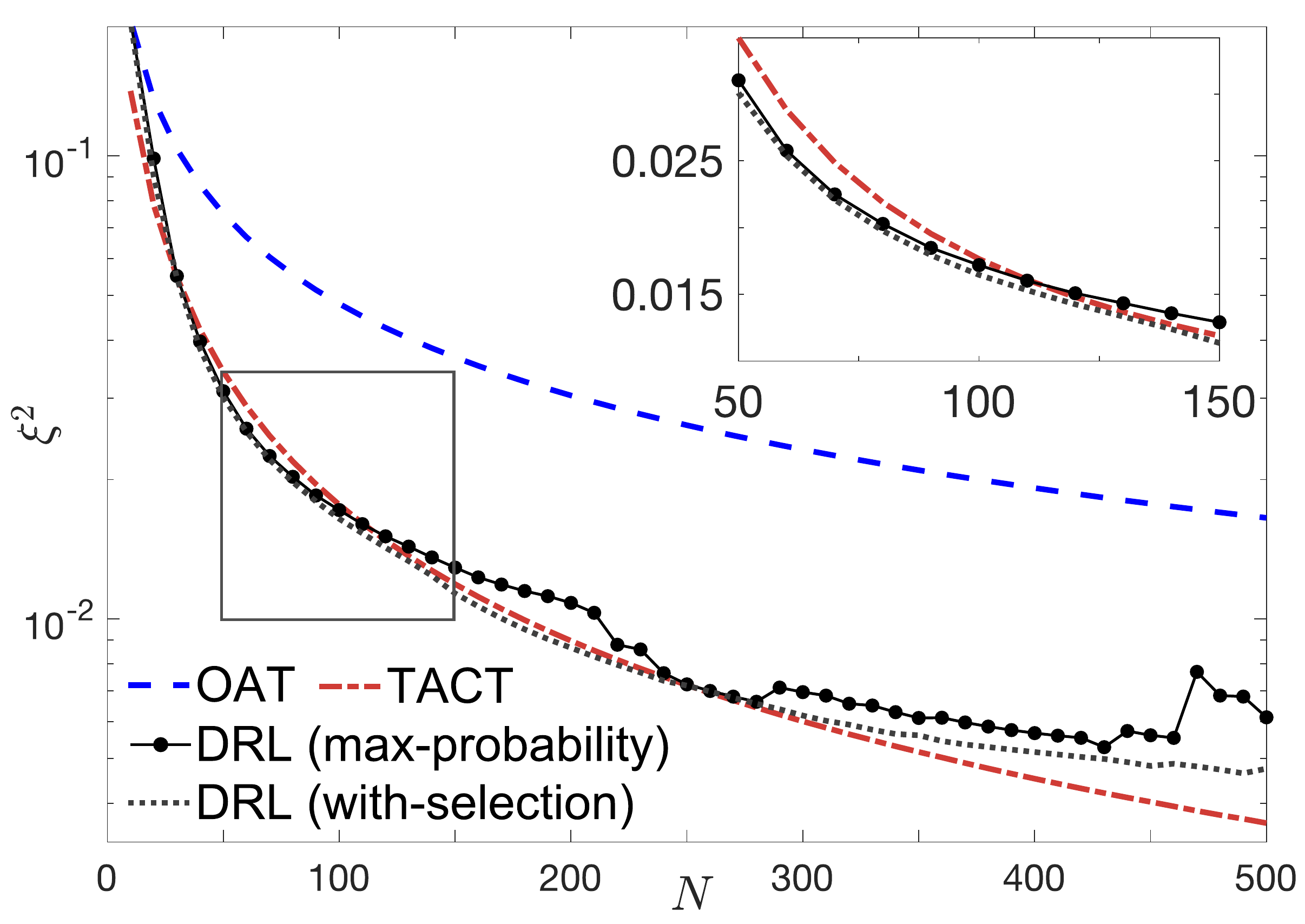}
\caption{
The $N$-dependence of the optimal SS coefficients obtained by applying the optimal DRL policy trained with $N = 100$ to other $N$ (black filled circles with solid guiding line and black dotted line). 
For comparison, the SS coefficients for OAT (blue dashed line) and the effective TACT (red dash-dotted line) are also shown.
Inset: zoomed-in view of the marked region with $N\in [50,150]$.
}
\vspace{-0.7cm}
\label{fig2}
\end{center}
\end{figure}

We next apply the optimal policy trained by DRL for $N = 100$ to systems with different particle numbers. 
The results are shown in Fig. \ref{fig2}. 
The optimal policy is found to perform well for $50 \leq N \leq 150$. 
However, for larger $N \gtrsim 300$, it fails to reach extreme SS (even for selected protocol). 
Therefore, the optimal policy found by DRL is size-dependent, and re-training is typically required for larger systems. 
Nevertheless, DRL process is relatively hard to train, especially for large-sized systems. 
For typical condensates containing $10^3\sim 10^5$ atoms or more, besides demanding for a large amount of computational resource, the increasingly larger Hilbert space dimension also makes it prohibitively expensive to explore the complete state space effectively.

Fortunately, from the results trained by DRL for small-sized systems, size-independent universal rules are discovered in the squeezing manipulation. 
First, we note $\theta$ and $-\theta$ pulses always appear in pairs, with $\theta=\pi/2$ giving the best performance. 
Other $\theta$ also works well and an example of $\theta=\pi/3$ is included in the Supplementary Material \cite{supp2}. 
Second, the first pulse pair appears at the moment when the effective TACT begins to outperform OAT. 
Third, the subsequent pulse pairs are applied whenever the current squeezing becomes inferior to that of OAT. 
These rules can be understood in physics as constituting a controlled rectification of over-twisting \cite{opatrny2015}. 
The bare OAT interaction twists the distribution of a spin state on the Bloch sphere around the $z$-axis nonlinearly, which is accompanied by a clockwise nonlinear rotation of the state around the $x$-axis.
This causes over-twisting of the state at later times, degrading the squeezing performance \cite{opatrny2015}. 
We note the combination of the $\pm\pi/2$ pulse pair found above can be described by the time evolution operator
\begin{equation}
U = e^{i\frac{\pi}{2}J_y}e^{-i\chi J_z^2\delta t}e^{-i\frac{\pi}{2}J_y} \approx e^{-i\chi J_x^2\delta t},
\label{pair_pulses}
\end{equation}
which corresponds to nothing but a nonlinear rotation operation along the $x$-axis~[see Fig.~\ref{fig1}(b)].
The $\pm\pi/2$ pulse pair therefore rectifies the nonlinear over-twisting around the $x$-axis, such that squeezing can be continuously improved until the state becomes over-twisted again. 
It is worthy to note that the scheme found by DRL is quite different from previous results of enhanced SS with controlled pulses. 
For instance, different from Ref.~\cite{liu2011}, the pulse sequence found here is nonperiodic, hence the system dynamics cannot be described by an effective TACT Hamiltonian. 
Furthermore, exact $\pm\pi/2$ pulses are required in Ref.~\cite{liu2011}, while the rectification of over-twisting here is independent of the specific choice of pulse area. For example, $\theta=\pi/3$ also works well as shown in the Supplementary Material \cite{supp2}.
In particular, the rectification we find here is nonlinear, which facilitates significantly enhanced SS within a short evolution time. 
This desirable feature is absent when using a transverse field along the mean spin direction \cite{duan2013,opatrny2015,jzhu2017,xiao2017} which only provides a linear complementary rectification. 
In the linear case, either greatly prolonged evolution time is required to obtain enhanced SS \cite{duan2013}, or efficient rectification to speed up SS is only maintained for a short time without improving the overall SS level \cite{opatrny2015,jzhu2017,xiao2017}.
 
\begin{figure}
\begin{center}
\includegraphics[width=\columnwidth]{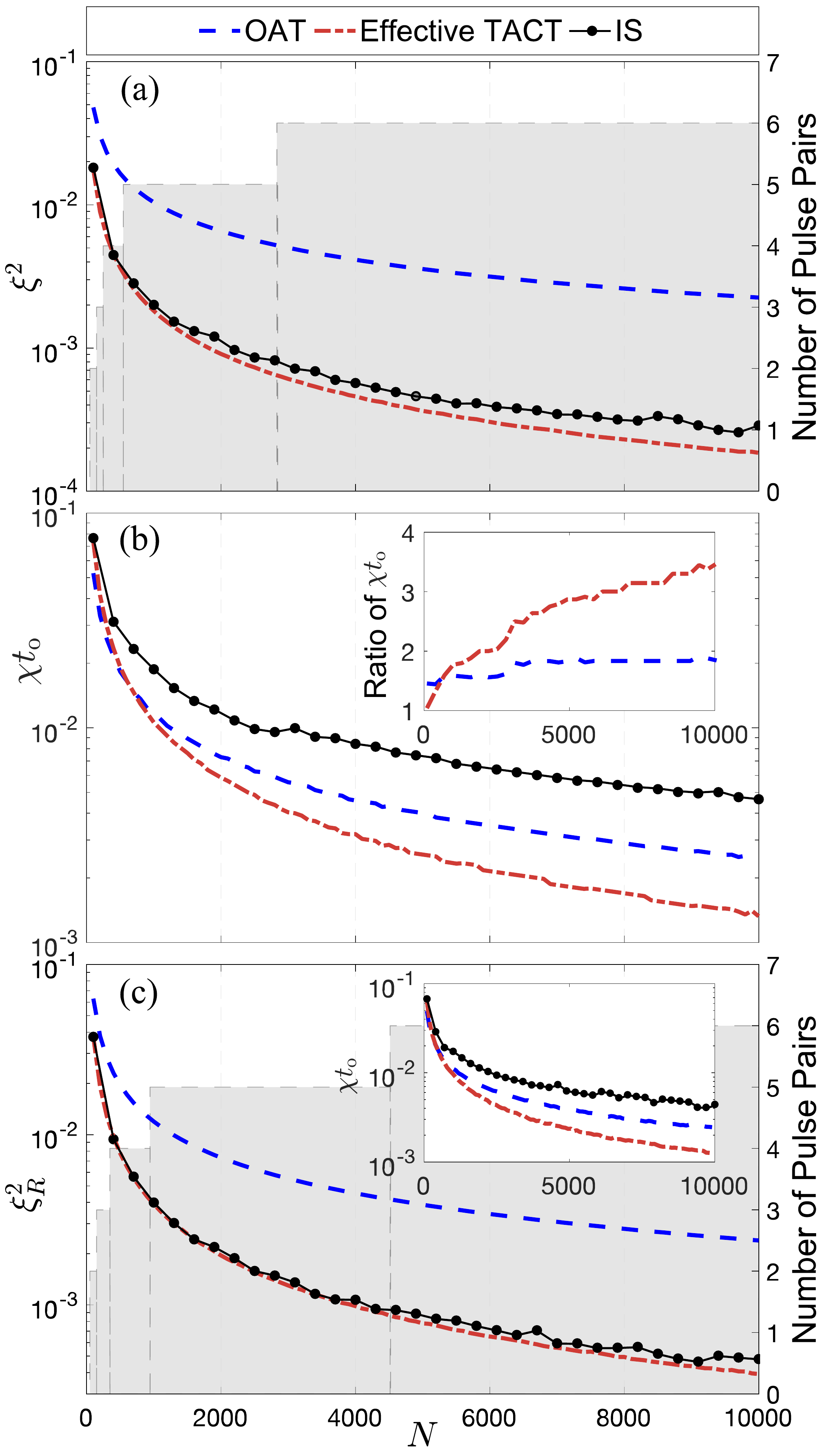}
\caption{The $N$-dependence of the optimal SS coefficients $\xi^2$ (a) and the corresponding evolution time (b) for the IS (black filled circles with solid guiding line), OAT (blue dashed line), and the effective TACT (red dash-dotted line).
Inset of (b): ratios of the optimal squeezing time of IS with respect to OAT ($t_{\rm{IS}}/t_{\rm{OAT}}$, blue dashed line),
and to the effective TACT ($t_{\rm{IS}}/t_{\rm{TACT}}$, red dash-dotted line). (c) Same as in (a) except that the vertical axis denotes the SS coefficients $\xi^2_R$.
Inset of (c): Same as in (b) except that the vertical axis denotes the corresponding evolution time when $\xi^2_R$ is optimal.
The shaded region in (a) and (c) displays the corresponding total number of pulse pairs used in IS (right vertical axes).
}
\vspace{-0.7cm}
\label{fig4}
\end{center}
\end{figure}

\begin{figure}
\begin{center}
\includegraphics[width=\columnwidth]{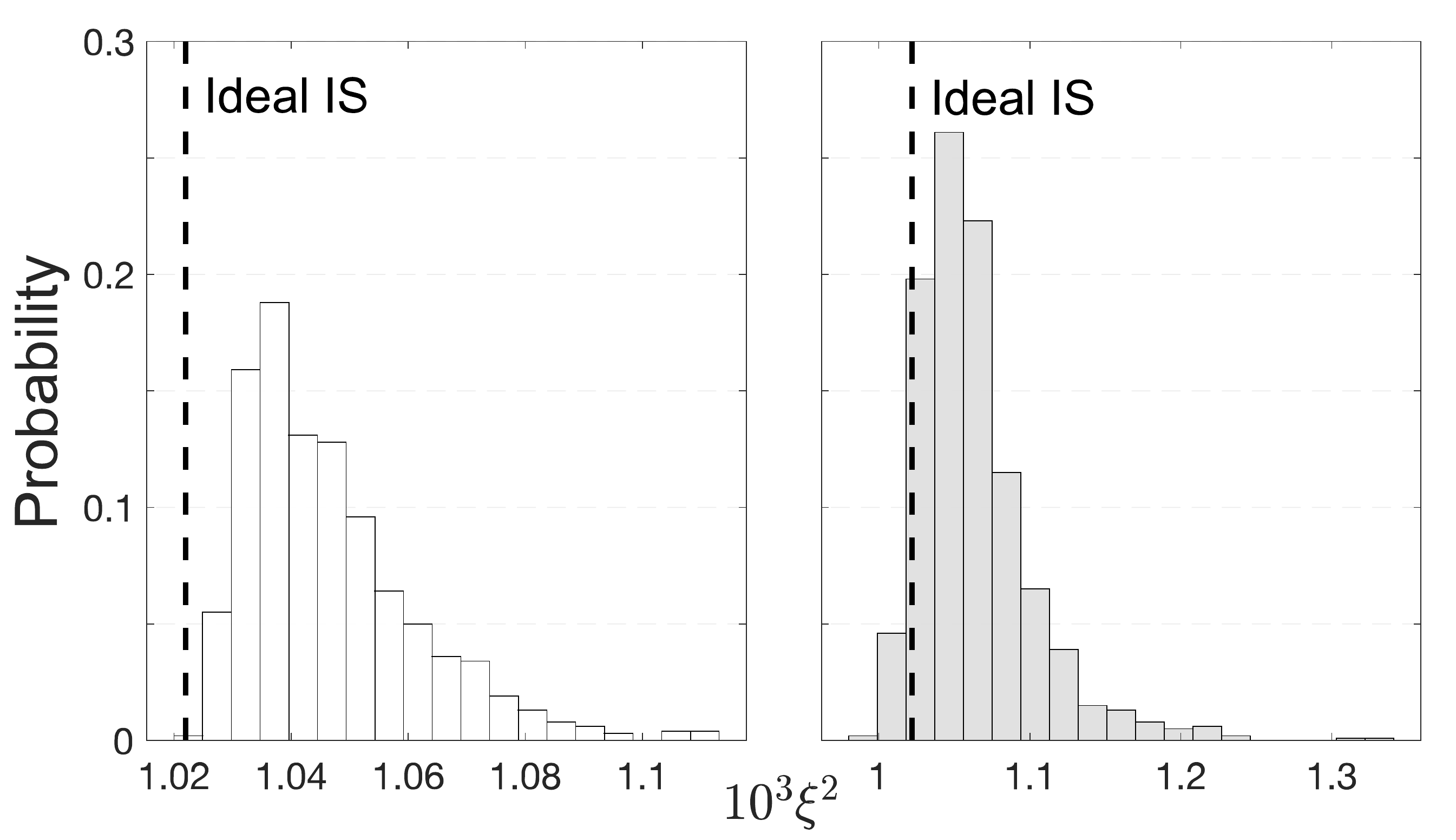}
\caption{Histograms from 1000 simulations with Gaussian stochastic distributions for $\delta t$ (left panel)
and temporal instants of pulse pairs (right panel) for $N = 2000$. The ideal IS without noise gives an optimal SS coefficient $\xi^2=1.02 \times 10^{-3}$ (black dashed line), while OAT gives $\xi^2 = 6.56 \times 10^{-3}$ and the effective TACT gives $\xi^2 = 0.91 \times 10^{-3}$.
}
\vspace{-0.7cm}
\label{fig5}
\end{center}
\end{figure}

Based on the above summarized universal rules and the insightful understanding gained, we next develop a DRL-inspired scheme (IS for short) applicable to large-sized systems yet demanding only for rather limited computation resource. 
Specifically, for a given $N$, we calculate the system dynamics and apply pulse sequences following the above three rules, with the duration $\delta t$ between the $\pm\pi/2$ pulses in a pulse pair further optimized by the grid search method \cite{supp2}.

In Fig. \ref{fig4}(a) we plot the optimal SS coefficient $\xi^2$ for systems with varying $N$. 
For IS, the optimal SS coefficient with respect to $N$ is found to scale as $\xi_{\rm{IS}}^2 \propto 1/N^{0.96}$, which is significantly better than the OAT case of $\xi_{\rm{OAT}}^2 \propto 1/N^{2/3}$, and approaches the extreme limit $\propto 1/N$. 
Moreover, the required number of pulse pairs increases rather slowly as $N$ becomes larger [gray shaded region in Fig. \ref{fig4}(a)]. 
For $N=10^4$, six pairs of pulses are essentially sufficient. 
In addition, the evolution time to achieve optimal SS remains on the same order of that for OAT or the effective TACT [see Fig. \ref{fig4}(b)]. 
The inverse power law scaling with respect to $N$ is found to be $\chi t_{\rm{IS}} \propto 1/N^{0.6}$, which is very close to that for OAT with $\chi t_{\rm{OAT}} \propto 1/N^{2/3}$.
We further illustrate the squeezing properties of our IS using an alternative definition of SS coefficient $\xi^2_R = N(\Delta\mathbf{J}_{\bot})_{\rm min}^2/\abs{\expval{\mathbf{J}}}^2$ with $\abs{\expval{\mathbf{J}}}$ being the mean spin length \cite{ramsey1, *ramsey1_2}, a coefficient which directly related to the measurement precision as it specifies the noise-to-signal ratio in Ramsey interferometry \cite{ramsey1, *ramsey1_2}.
From Fig. \ref{fig4}(c), we see that the performance of the IS is even better with this squeezing definition. 
The fitted power law scales as $\xi_{\rm{R,IS}}^2 \propto 1/N^{0.98}$, and $\chi t_{\rm{R,IS}} \propto 1/N^{0.6}$.

Since only a small number of pulses are required, the IS can be potentially robust against noise or stochastic error.
In actual experiments, the dominant source of error often comes from fluctuations in temporal instants of the pulse pairs or in the duration $\delta t$ within each pulse pair.
We perform numerical simulations by modeling these two sorts of errors as Gaussian stochastic processes.
The consequent optimal squeezing coefficient $\xi^2$ under 1\% noise level is shown in Fig. \ref{fig5}, which clearly implicates that the IS is rather robust against the aforementioned noises. 
In a previously reported OAT experiment \cite{gross2010} involving two hyperfine states $\ket{1,1}$ and $\ket{2,-1}$ of $^{87}\rm{Rb}$ atoms, atomic coupling strength reaches $\chi = 2\pi \times 0.063\, \rm{Hz}$. At $N =2000$, a $1\%$ uncertainty in time duration amounts to $\delta t \sim 0.4\, \mu\rm{s}$, while a $1\%$ uncertainty in the temporal instant of the pulse corresponds to $\sim 0.3\, \rm{ms}$.
Both limits are easily satisfied within current experimental capabilities.

Particle loss or detection noise in real experiments also degrade the level of achievable squeezing, as they are known to prevent from reaching the OAT limit \cite{gross2010,riedel2010, liyun2008}.
The potential limitations of these factors on our protocol are expected to be on the same level as in reported OAT experiments, due primarily to the comparably required time duration to reach optimal SS.
Further improvements can come from continued development of experimental techniques \cite{gross2010,riedel2010,hosten2016,haine2018}.
Once the OAT limit is achieved, the advantages of our scheme can be readily demonstrated.

In conclusion, we propose to realize extreme SS based on OAT using a few control pulses.
Our scheme is based on size-independent universal rules for squeezing manipulation, discovered with the assistance of DRL.
They can be understood as facilitating nonlinear rectification of over-twisting associated with OAT continuously, hence precisely direct the evolution to optimal SS.
Based on our IS, realizable SS coefficient is found to scale as $\propto 1/N^{0.98}$ significantly surpassing the bare OAT limit of $\propto 1/N^{2/3}$, and approaches the extreme SS limit $\propto 1/N$. 
The controlled manipulations required are modest, composing of only a few rotation pulses applied at appropriate instants. For example, 6 pairs of pulses are sufficient for a system with up to $10^4$ particles.
Moreover, the time to realize optimal SS is found to remain on the same order of the optimal OAT squeezing time, which indicates that our scheme can be as robust as OAT with respective to particle loss or decoherence.
In our scheme, DRL plays a crucial role in providing insights for discovering universal rules, which paves the way for an experimentally feasible scheme within the state-of-the-art lab techniques.
Our work highlights the great potential of applying DRL to controlled quantum dynamics and quantum state engineering.

\begin{acknowledgments}
We thank Q. Liu for fruitful discussions.
This work is supported by the National Key R\&D Program of China (Grant No. 2018YFA0306504)
and the NSFC (Grant No. 91636213, No. 11654001, and No. 91736311).
Y.C. Liu acknowledges support from the NSFC (Grant No. 91736106,  11674390, 91836302).
L. You also acknowledges support from BAQIS (Grant No. 20191550064).
\end{acknowledgments}


%

\clearpage
\onecolumngrid
\begin{center}
  \textbf{\large Supplementary Material}\\[.2cm]
\end{center}

This supplementary provides discussions that were left out of the main text due to space limitations. First, we briefly introduce the key concepts of deep reinforcement learning (DRL) and specify our optimization and search problem.
Then, we will give the results of a DRL task with $\pi/3$ pulses.
This is followed by introduction of grid search optimization for IS and comparative studies of squeezing coefficient between DRL policy and DRL-inspired scheme (IS).
In the end, we discuss the robustness of our policy against noise of the control pulses.

\section{deep reinforcement learning task}
\begin{figure}[!htp]
\begin{center}
\includegraphics[width=0.6\textwidth]{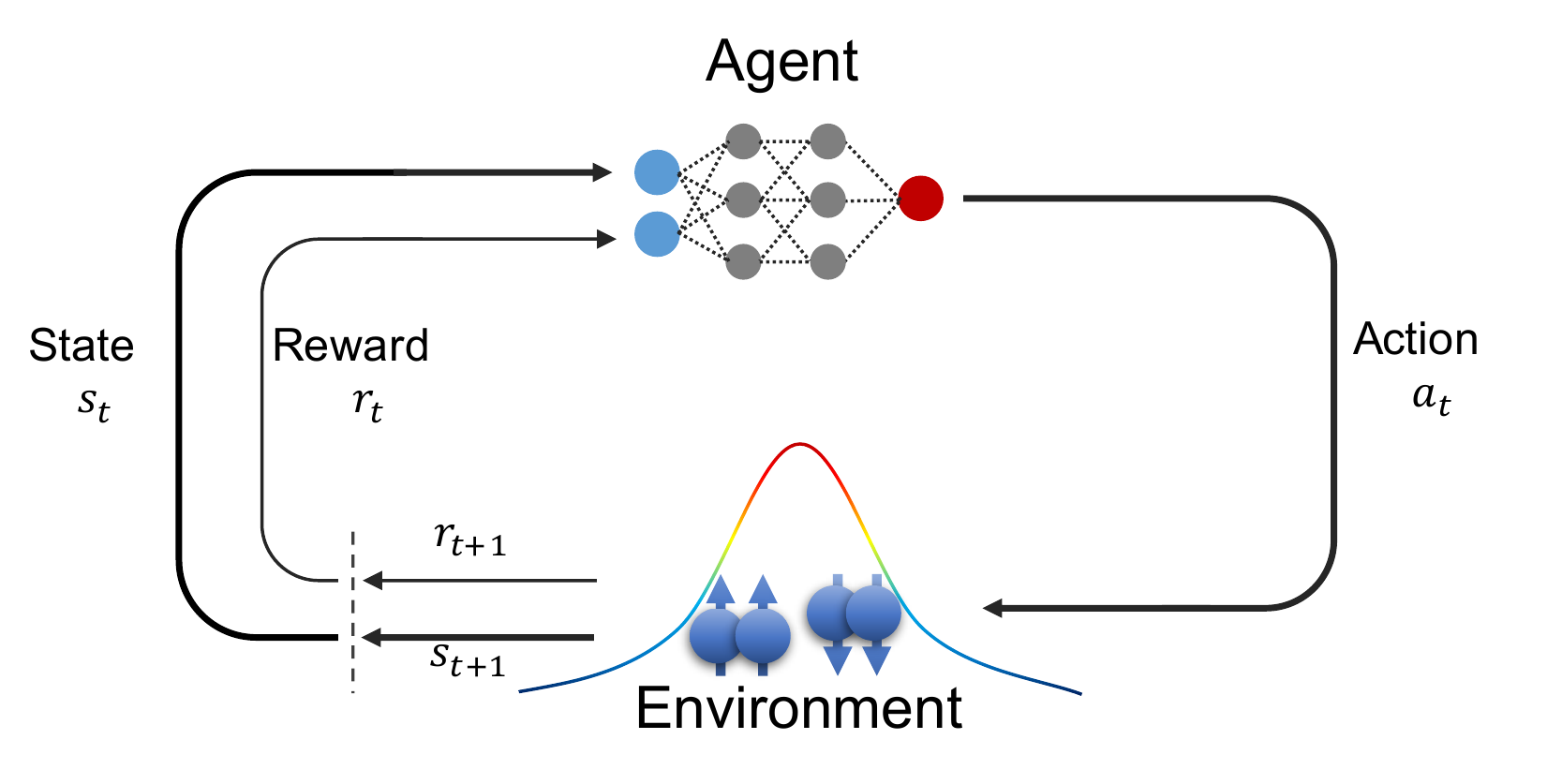}
\caption{Illustration of a typical Markov decision process (MDP) in DRL.}
\label{fig_sup1}
\end{center}
\end{figure}

DRL task is modeled by a Markov decision process (MDP) as shown in Fig. \ref{fig_sup1}. At each time step $t$, the agent obtains the observation $s_t \in \mathcal{S}$ of the environment and select an action $a_t \in\mathcal{A}$ based on current policy $\varpi(a_t|s_t)$. The environment then evolves to $s_{t+1}$ due to action $a_t$ and returns a scalar reward $r_t\in\mathcal{R}$ back to the agent. Policy $\varpi$ is consequently updated through such experiences data in some way to maximize accumulated rewards.

The definitions of state $\mathcal{S}$, action $\mathcal{A}$, and reward function $\mathcal{R}$ in the spin squeezing problem are as follows.

\begin{itemize}
\item $\mathcal{S}$: For the collective spin system, we choose $\{\expval{\mathbf{J}},\expval{J_x^2},\expval{J_z^2}\}$ as the state representation instead of its wave function $\ket{\psi}$, although the latter contains complete information. Using such relevant physical observables as state representation makes
    the policy directly generalizable and the final policy easily interpretable with clear physical insights. Here, the mean spin $\expval{\mathbf{J}}=(\expval{J_x},\expval{J_y},\expval{J_z})$ provides the direction of the collective spin and $\{\expval{J_x^2}, \expval{J_z^2}\}$
    contains correlations between different atomic spins.

\item $\mathcal{A}$: The action space consists of three discrete operations $\{0,\pm\pi/2\}$, where $a=0$ means evolving for a time interval $\delta t$ under bare OAT Hamiltonian $H_{\rm{OAT}} = \chi J_z^2$, while $a=\pi/2(-\pi/2)$ means the same action $0$ preceded by a $\pi/2(-\pi/2)$ pulse.

\item $\mathcal{R}$: We use squeezing coefficient $\xi^2=4(\Delta\mathbf{J}_{\bot})_{\rm{min}}^2/N$ as the object function to optimize for spin squeezing \cite{ueda1993}. The total achievable reward is $R_{\rm{tot}}=\xi^2_i - \xi^2_f$, where $\xi^2_i = 1$ for our chosen initial CSS and $\xi_f^2$ is the squeezing ratio at final time.
    To avoid the problem of sparse rewarding, we decompose the object function into a summation form:
\begin{equation}
\label{ }
R_{\rm{tot}}=\sum_{j=1}^n\xi^2(t_{j-1}) - \xi^2(t_j).
\end{equation}
At each time step $t_j$, a reward $r_j = \xi^2(t_{j-1}) - \xi^2(t_j)$, the instantaneous decrease of spin squeezing coefficient, is fed back to the agent. Dense rewards make the training process quicker and more stable. Furthermore, the reward can be modified into
\begin{equation}
\label{ }
r_j \rightarrow \log_{10} ( \xi^2(t_{j-1})/\xi^2(t_{j}) ),
\end{equation}
which prevents the decline of learning efficiency when squeezing approaches to the minimum.
\end{itemize}

The proximal policy optimization (PPO) algorithm is employed to find the optimized policy $\varpi^*$ that can maximize the cumulative reward $R$,
\begin{equation}
\label{ }
\varpi^* = \mathop{\arg\max}_{\varpi} R, ~~\mathrm{with} ~~ R=\sum_j\gamma^tr_j,
\end{equation}
where $\gamma$ is a discount factor. It is typically chosen very close to 1 to avoid greedy solutions. The pseudo-code of PPO algorithm is shown in Table \ref{PPO_pseudo_code} \cite{ppo}. To facilitate the training process, we encapsulated the evolution of spin state into a gym environment as suggested by openAI \cite{gym}. In PPO algorithm, the policy is stochastic which returns a distribution on action space for a given state $s$. Here, in our problem on optimizing spin squeezing, the action space $\mathcal{A}$ contains only three operations, so the policy reduces to a discrete distribution that satisfies
\begin{eqnarray}
\sum_{a\in\{ 0,\pm\pi/2 \}}\varpi(a|s) \equiv 1.
\end{eqnarray}

Once a policy is learnt, we can either choose the action with maximum probability as $a=\max_{a'}\varpi^*(a'|s)$ to give a deterministic protocol or obtain a protocol by selecting the best one among multiple protocols from a simple sampling based on $\varpi^*(a'|s)$.

For a system with total particle number $N=100$, a simple neural network
is used to parameterize the actor and critic network in PPO.
Both of them contain two fully connected hidden layers with $32$ and $16$ neurons respectively.
Each learning episode is divided into 50 consecutive steps and the total evolution time is limited to $\chi t_c=0.08$. The other hyper-parameters used in our training are listed in Table \ref{table-ss-1}. No sophisticated hyper-parameter tuning is used in our problem.

\begin{table}[!htp]
\centering
\caption{Pseudo-code of PPO algorithm}
\begin{tabular}{l}
\hline
PPO algorithm \\
\hline
1. Input: initial weights of policy network $\theta_0$, initial weights of value function $\phi_0$ \\
2. $\bold{for}$ k = 0, 1, 2, ... $\bold{do}$ \\
3. ~~~~~~Collect multiple trajectories of spin state evolution $\mathcal{D}_k=\{\tau_i\}$ under current policy $\varpi_{\theta_k}$ \\
~~~~~~~~~~in self-defined Gym environment. \\
4. ~~~~~~Computes rewards-to-go $\hat{R}_t=\sum_{l=0}^{\chi t_c-t-1}\gamma^l r_{t+l}+V_{\phi_k}(s_{\chi t_c})$. \\
5. ~~~~~~Use GAE-$\lambda$ method and current value function $V_{\phi_k}$ to estimate advantage function $A^{\varpi_k}$. \\
6. ~~~~~~Maximize PPO-Clip lower bound function and update weights of policy network \\
~~~~~~~~~\qquad\qquad\qquad $\displaystyle \theta_{k+1}=\mathop{\arg\max}_\theta \dfrac{1}{|\mathcal{D}_k|\chi t_c}\sum_{\tau\in\mathcal{D}_k}\sum_{t=0}^{\chi t_c} L_k^{\text{clip}}(s_t,a_t)$, \\
~~~~~~~~~~usually using gradient descents methods such as Adam and SGD. \\
7. ~~~~~~Minimize mean-squared error and update weights of value function \\
~~~~~~~~~\qquad\qquad\qquad $\displaystyle \phi_{k+1}=\mathop{\arg\min}_{\phi} \frac{1}{|\mathcal{D}_k|\chi t_c}\sum_{\tau\in\mathcal{D}_k}\sum_{t=0}^{\chi t_c} \left( V_{\phi}(s_t)-\hat{R}_t \right)^2$, \\
~~~~~~~~~~usually using gradient descents methods such as Adam and SGD. \\
8. $\bold{end~for}$ \\
\hline
\end{tabular}
\label{PPO_pseudo_code}
\end{table}

\begin{table}[!htp]
\caption{Training Hyperparameters for PPO}
\centering
\begin{tabular}{rl}
\hline
Hyperparameters& \hskip 6pt value\\
\hline
hidden size& \hskip 6pt [32, 16] \\
activation& \hskip 6pt $\tanh$ \\
discounted factor $\gamma$& \hskip 6pt 0.9 \\
actor-network learning rate& \hskip 6pt 3E-4 \\
critic-network learning rate& \hskip 6pt 1E-3 \\
target KL-divergence& \hskip 6pt 0.01 \\
clip ratio $\epsilon$& \hskip 6pt 0.2  \\
GAE-$\lambda$& \hskip 6pt 0.97  \\
\hline
\end{tabular}
\label{table-ss-1}
\end{table}

\section{DRL protocol with $\pi/3$ pulse for N = 100}
\begin{figure}[!htp]
\begin{center}
\includegraphics[width=0.6\textwidth]{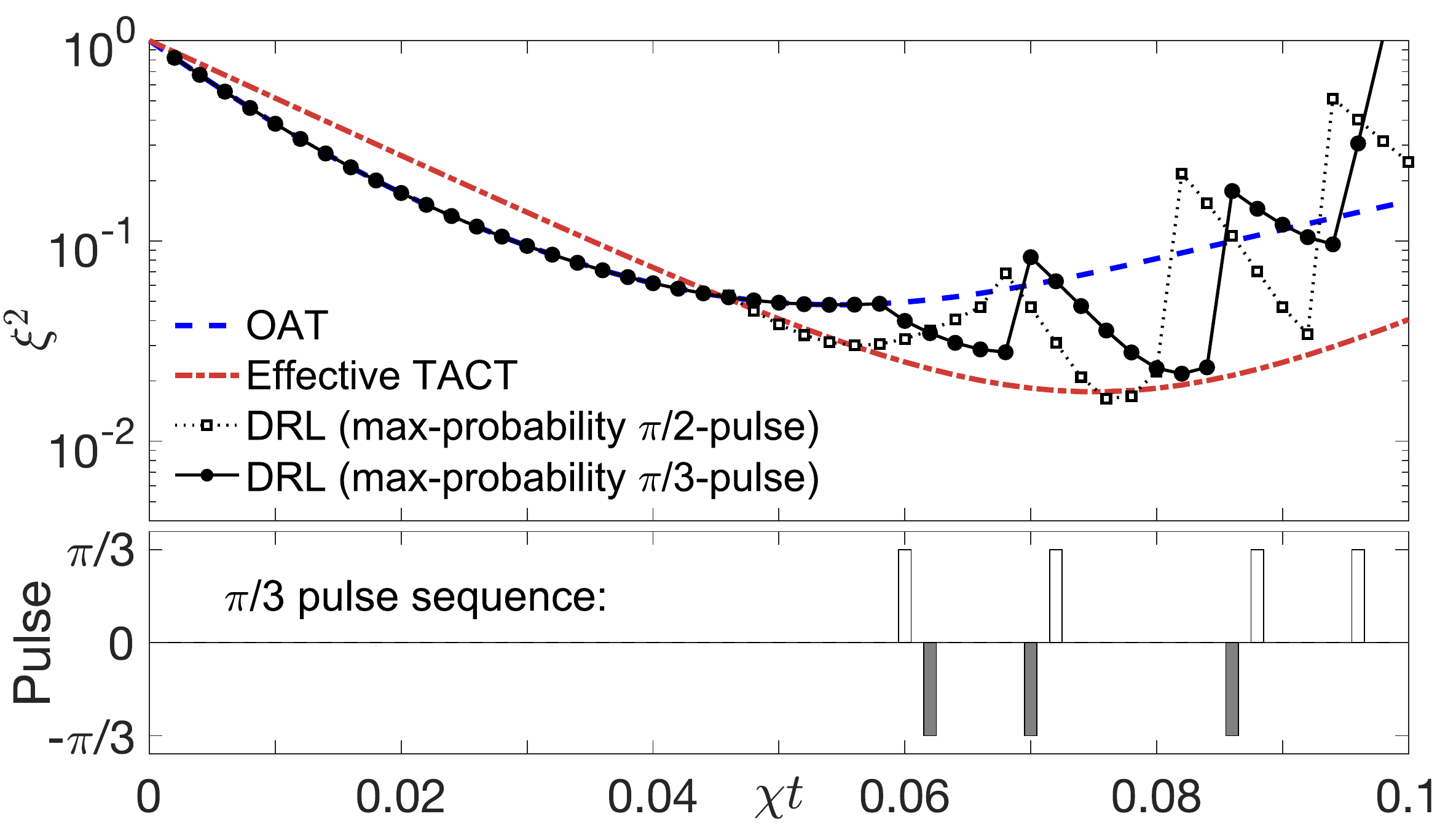}
\caption{Evolution of spin squeezing coefficients are compared among the max-probability given by DRL policy with $\pi/2$ pulses (open squares with black dotted guiding line) and $\pi/3$ pulses (black filled circles with solid guiding line), OAT (blue dashed line) with Hamiltonian $H_{\rm{OAT}}=\chi J_z^2$ and the effective TACT (red dashed-dotted line) with $H_{\rm{TACT}} = \chi(J_z^2 - J_y^2)/3$ for $N = 100$. The initial state is a CSS along the $x$-axis. The lower panel illustrates the corresponding DRL pulse sequence that forms the agent's most significant operations: $(\pi/3,-\pi/3)$ and $(-\pi/3,\pi/3)$ pulse pairs in the $\pi/3$-pulse sequence.} 
\label{fig_sup1_5}
\end{center}
\end{figure}

In this section, we illustrate the underlying mechanism of the DRL policy in improving SS by using another form of pulse, with $\int^{t_0 + \Delta t}_{t_0} \Omega(t)dt = \pi/3$. We call it 
$\pi/3$-policy in the following. In the $\pi/3$-pulse training task, all the hyper-parameters are chosen as same as the case of $\pi/2$-pulse (Table.\ref{table-ss-1}).
The results for a system with $N=100$ are shown in Fig. \ref{fig_sup1_5}.
By comparing the pulse sequences suggested by DRL policy in the $\pi/2$-policy (discussed in the main text) and the $\pi/3$-policy, we find that they share similar features: pulses come in pair and are applied around the instants when the effective TACT outperforms OAT or when the current squeezing is lagging behind that of OAT.
%
For the $\pi/3$ case, the time evolution operator $U$ for a $(\pm\pi/3,\mp\pi/3)$ pulse pair reads,
\begin{equation}
\label{ }
U = e^{\pm i\frac{\pi}{3}J_y}e^{-i\chi J_z^2\delta t}e^{\mp i\frac{\pi}{3}J_y} = e^{-i\chi(J_x\sin(\frac{\pi}{3}) \pm J_z\cos(\frac{\pi}{3}))^2\delta t}
\simeq e^{-i\frac{3}{4}\chi J_x^2\delta t}e^{\mp i\frac{\sqrt{3}}{4}\chi(J_xJ_z + J_zJ_x)\delta t}e^{-i\frac{1}{4}\chi J_z^2\delta t},
\end{equation}
where we have neglected the high-order $\mathcal{O}(\delta t^2)$ term in the last equality.
Again, we find the non-linear term $\propto J_x^2$, which contributes to the rectification of the over-twisting problem as discussed in the main text.  Note that here we have additional crossing term $J_xJ_z+J_zJ_x$, which is absent in the $\pi/2$-pulse case and degrades the rectification effect. It can be partly eliminated by alternatively and consecutively applying $(\pi/3,-\pi/3)$ and $(-\pi/3,\pi/3)$, whose crossing terms have opposite signs. 

\section{Grid Search for IS and Comparison between DRL and IS}
\begin{figure}[!htp]
\begin{center}
\includegraphics[width=0.985\textwidth]{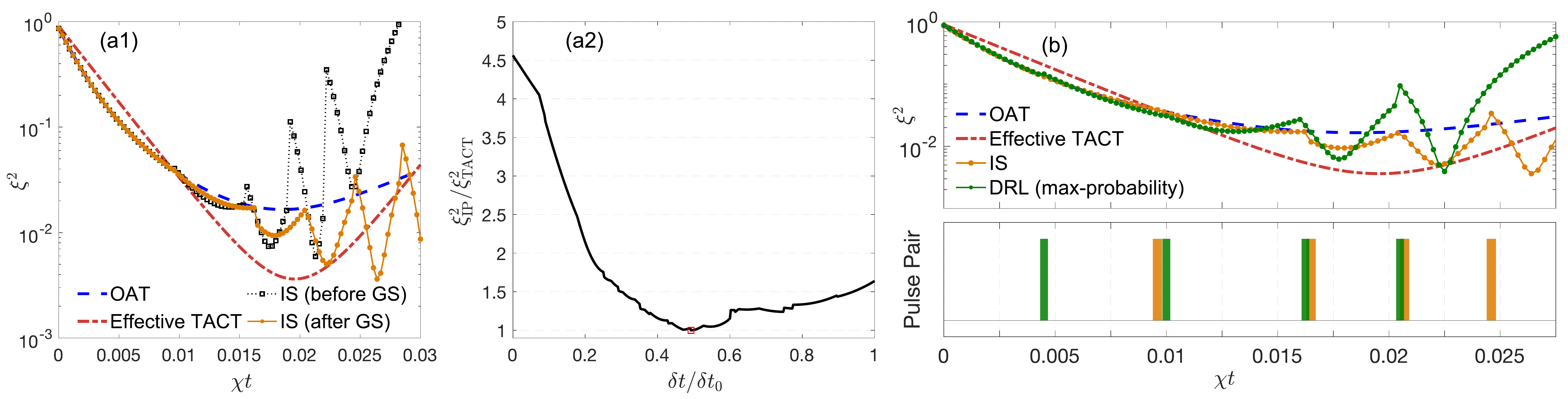}
\caption{(a1) Evolution of spin squeezing coefficients are compared among IS (before GS) (open squares with black dotted guiding line), IS (after GS) (orange filled circles with solid guiding line), OAT (blue dash line) and effective TACT (red dash-dotted line) at $N = 500$. (a2) Grid searching results of $\delta t$ in $[0,\delta t_0]$ region and the red open square mark denotes the optimal position corresponding to IS (after GS) in (a1).
(b) Evolution of the squeezing coefficients from various schemes: OAT, the effective TACT, DRL (max-probability) (green filled circles with solid guiding line), and IS (orange filled circles with solid guiding line) for $N = 500$. The corresponding pulse pair sequences for DRL (green bars) and IS (orange bars) are shown in the lower panel.}
\label{fig_sup2}
\end{center}
\end{figure}

In our IS, the whole evolution is also divided into $M$ steps with equal duration $\delta t_0$. Initially, the time duration $\delta t$ between $\pm\pi/2$ pulses in a pulse pair is set as $\delta t_0$ and the pulse pair applying moment is determined by our concluded rules from DRL protocols. However, the direct implement of IS in a $N=500$ system (open squares with black dotted guiding line in Fig. \ref{fig_sup2}(a1)) cannot reach our expectation, performing as well as DRL policy (green filled circles with solid guiding line in Fig. \ref{fig_sup2}(b)), since the DRL policy can fine tune the pulse applying moments around the instants we concluded. We noticed that the time duration $\delta t$ in a pulse pair directly determines the non-linear rectification degree on OAT induced over-twisting state, such that the performance of our IS can be enhanced via optimizing $\delta t$. This optimization process is done through a so-called grid search (GS) calculation. That is, we scan the $\delta t \in [0,\delta t_0]$ and figure out an optimal $\delta t$ leading to the best performance. The searching process is shown in Fig. \ref{fig_sup2}(a2) and the optimized IS result is compared with OAT and effective TACT in Fig. \ref{fig_sup2}(a1). After a grid search optimization, the squeezing performance of our IS is approaching to TACT.

To demonstrate that we have captured the key characteristic features of the DRL policy, we first compare the squeezing coefficients obtained with DRL and IS for $N = 500$ as shown in Fig. \ref{fig_sup2}(b). For the DRL policy, the maximum number of pulse pairs is limited to 4 and the action space $\mathcal{A}$ is contracted to $\{0,1\}$, where `1' means applying a pulse pair $(\pi/2,-\pi/2)$ and `0' means free evolution under bare OAT Hamiltonian for $\delta t$.

The optimal squeezing for both policies reach the same extreme level of the effective TACT, with DRL approaching the optimal squeezing at an earlier moment. Taking the computation effort into account, IS is obviously favored. The lower panel of Fig. \ref{fig_sup2} shows the corresponding pulse sequence.
Out of four pulse pairs, three are found to be applied essentially at the same moments for these two policies.

\section{Robustness against fluctuations in control pulses}
\begin{figure}[!htp]
\begin{center}
\includegraphics[width=0.985\textwidth]{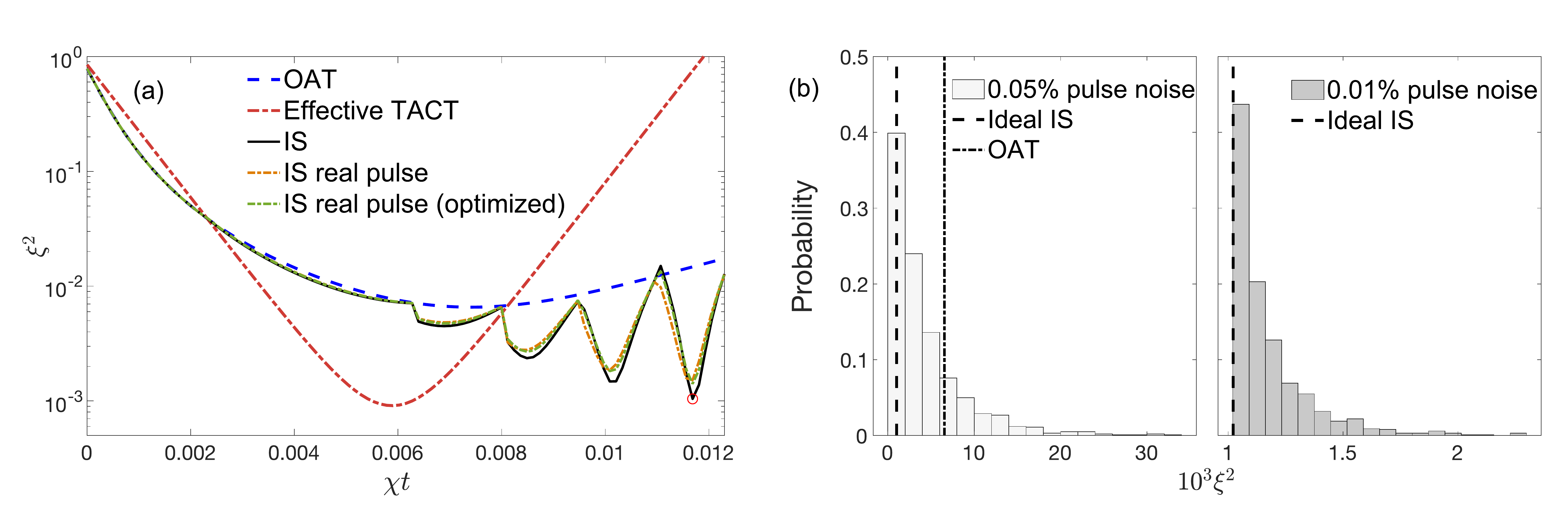}
\caption{Spin squeezing in the presence of real pulses or with noise in pulse area at $N = 2000$. (a) Comparisons among OAT, the effective TACT, IS, IS for real pulses with ideal or optimal pulse sequence, and IS for real pulses with a re-optimized $\delta t$ (or sequence). (b) A histogram of the squeezing coefficient for 1000 simulation samples at $\chi t = 0.0117$ (the red open circle in (a)) with a $0.05\%$ ($0.01\%$) Gaussian stochastic uncertainties in pulse area for the left (right) panel. The ideal IS gives an optimal squeezing coefficient $\xi^2 = 1.02 \times 10^{-3}$ (black dashed lines), while the OAT gives $\xi^2 = 6.6 \times 10^{-3}$ (black dash-dotted line).}
\label{fig_sup3}
\end{center}
\end{figure}

In previous discussion, our numerical calculations for IS always assume that the $\pm\pi/2$ control pulses are executed
so quickly that during it the OAT interaction is negligible.
Typically, for a $10$ kHz microwave Rabi frequency, the duration of a $\pi/2$ pulse is about 25 $\mu$s, which translates into about $10\%$ of the time duration $\delta t$ between pulse for $N = 2000$ and $\chi = 2\pi \times 0.063$ Hz in $^{87}\rm{Rb}$ \cite{gross2010}. In Fig. \ref{fig_sup3}(a), phase accumulation from the bare OAT Hamiltonian over the applied pulse is considered. It will lead to a degradation of the achievable spin squeezing (the orange dash-dotted line in Fig. \ref{fig_sup3}(a)), although some of the squeezing performance can be rectified after a $\delta t$ optimization with respect to real pulse operation (the green dash-dotted line in Fig. \ref{fig_sup3}(a)).

Next, we consider imperfections of the pulse area for a square-shaped versus a Gaussian-shaped pulse we used. Due to the associated fluctuation phase mismatch of the $\pm\pi/2$ pulse in a pair, the mean spin will be rotated along the $y$-axis, i.e. $\expval{\mathbf{J}} = (\expval{J_x},0,0) \rightarrow (\expval{J_x}\cos(\theta_r),0,\expval{J_x}\sin(\theta_r))$, where $\theta_r$ is the net non-ideal cumulative phase during a pulse pair.
The subsequent squeezing of such a state is dramatically degraded under bare OAT Hamiltonian evolution. We have numerically simulated
such imperfections using $\pi/2$ pulses with 0.05\% and 0.01\% Gaussian stochastic uncertainties and the squeezing coefficient
is estimated assuming that
 mean spin still points along the $x$-axis $\expval{\mathbf{J}} = (\expval{J_x},0,0)$.
The squeezing coefficient $\xi^2$ shown in Fig. \ref{fig_sup3}(b) is sampled at $\chi t = 0.0117$,
corresponding to the optimal squeezing position of our IS at $N =2000$ system (the red open circle mark in Fig. \ref{fig_sup3}(a)).
The squeezing coefficient $\xi^2$ under such a fluctuating error distribution to the $\pm\pi/2$
pulse area in most of the samples is found to still outperform that of OAT.

\end{document}